\documentclass{aastex631}

\begin{document}

\title{The Kinematical Behavior of Solar Eruptive Filaments Affected by the Poloidal Magnetic Field}

\correspondingauthor{Ye Qiu}
\email{qiuye@smail.nju.edu.cn}

\correspondingauthor{Zhen Li}
\email{lizhen@nju.edu.cn}

\author[0000-0002-1190-0173]{Ye Qiu}
\affiliation{Institute of Science and Technology for Deep Space Exploration, Suzhou Campus, Nanjing University, Suzhou 215163, China}

\author[0000-0002-9293-8439]{Yang Guo}
\affiliation{School of Astronomy and Space Science, Nanjing University, Nanjing 210023, China }
\affiliation{ Key Laboratory of Modern Astronomy and Astrophysics (Nanjing University), Ministry of Education, Nanjing 210023, China}

\author[0000-0002-4978-4972]{Mingde Ding}
\affiliation{School of Astronomy and Space Science, Nanjing University, Nanjing 210023, China }
\affiliation{ Key Laboratory of Modern Astronomy and Astrophysics (Nanjing University), Ministry of Education, Nanjing 210023, China}

\author[0000-0001-7693-4908]{Chuan Li}
\affiliation{Institute of Science and Technology for Deep Space Exploration, Suzhou Campus, Nanjing University, Suzhou 215163, China}
\affiliation{School of Astronomy and Space Science, Nanjing University, Nanjing 210023, China }
\affiliation{ Key Laboratory of Modern Astronomy and Astrophysics (Nanjing University), Ministry of Education, Nanjing 210023, China}

\author[0000-0003-1350-9722]{Linggao Kong}
\affiliation{Institute of Science and Technology for Deep Space Exploration, Suzhou Campus, Nanjing University, Suzhou 215163, China}

\author{Zhen Li}
\affiliation{School of Astronomy and Space Science, Nanjing University, Nanjing 210023, China }
\affiliation{ Key Laboratory of Modern Astronomy and Astrophysics (Nanjing University), Ministry of Education, Nanjing 210023, China}

\begin{abstract}
Kinematics of solar eruptive filaments is one of the important diagnostic parameters for predicting whether solar eruptions would induce geomagnetic storms. Particularly, some geomagnetic storms might be induced by solar filament eruptions originating from unexpected surface source regions because of non-radial ejection. The non-radial ejection of filaments has received widespread attention but remains inconclusive. We select two eruptive filaments, both of which are supported by flux ropes, as indicated by the hot channel structures seen in the 94 Å images and the hook-shaped brightenings where the filament material falls back. We measure the three-dimensional ejection trajectory of the eruptive filaments by integrating the simultaneous observations from SDO and STEREO. Furthermore, we calculate the distribution of the poloidal field along the ejection path and compare it to the ejection acceleration. It is revealed that the reinforcement of the poloidal magnetic field may lead to the suppression of the acceleration, with the acceleration resuming its increase only when the poloidal field diminishes to a certain level. Additionally, we compute the spatial distribution of the poloidal field in various directions and find that the poloidal magnetic field above the filaments is asymmetric. For both investigated events, the filaments appear to eject towards the side where the poloidal magnetic field is weaker, indicating that the eruptive filaments tend to propagate along the side with weaker strapping force. This may provide a new explanation for the inclined ejection of filaments.

\end{abstract}

\keywords{Solar filaments (1495) --- Solar filament eruptions (1981) --- Solar magnetic fields (1503) --- Solar coronal mass ejections (310) --- Solar flares (1496) }

\section{Introduction} \label{sec:intro}
Solar filament eruptions are one of the common activities occurring in the solar atmosphere. The eruptive filaments can evolve into coronal mass ejections (CMEs)
 when they propagate into the interplanetary space \citep{2003ApJ...586..562G,2012LRSP....9....3W}. Thus, filament eruptions can be regarded as a source of space weather, which may induce geomagnetic storms and cause huge damages to artificial high-tech facilities \citep{1981JGR....86.4555J,1993JGR....9818937G}. Even though geomagnetic storms are more likely driven by the front-side eruptive filaments, there are still many storms triggered by lateral eruptive filaments originating from the solar limb due to the deflection during ejection \citep{2003ApJ...582..520Z, 2007JGRA..11210102Z}. Therefore, it is one of the key issues to understand the causes of non-radial filament ejections for improving the prediction abilities of disastrous space weather.

The non-radial ejection of filaments is a general phenomenon, which can be reconstructed by multipoint observations \citep{2011ApJ...730..104J,2020ApJ...897...35R} and various modelling tools (e.g. \citealt{2009SoPh..256..111T,2016ApJ...833..267I,2021A&A...653L...2Z}). Certainly, the inversion of the spectroscopic observations is also an effective method to acquire the three-dimensional kinematics of eruptive filaments, e.g. as shown by \cite{2024ApJ...961L..30Q}, who utilized the full-disk high-spectral-resolution H$\alpha$ spectra supplied by the Chinese H$\alpha$ Solar Explorer (CHASE; \citealt{2022SCPMA..6589602L,2022SCPMA..6589603Q}). Multiple mechanisms explain why some filaments propagate in an inclined direction, but each has specific requirements for the magnetic configuration. One potential mechanism for the non-radial ejection of filaments is the magnetic reconnection that should either occur laterally to the filament structure \citep{2000ApJ...545..524C, 2022ApJ...933..148C} or generate horizontally bent field lines \citep{2013ApJ...779..129K}. Both scenarios intrinsically lead to an asymmetric distribution of magnetic tension forces. Additionally, active regions may trigger inclined filament eruptions provided that their vertically oriented opposite-polarity magnetic fields differ in strength by a factor of 2 or more \citep{2010ApJ...708..314A}. The non-radial ejection might also be dependent on magnetic pressure gradient caused by the existence of large-scale structures, such as coronal holes \citep{2009JGRA..114.0A22G,2013SoPh..287..391P,2023ApJ...953..150S}, heliospheric current sheets \citep{2015SoPh..290.3343L}, helmet streamers \citep{2018ApJ...862...86Y} and pseudostreamers \citep{2012ApJ...744...66Z,2020AdSpR..65.1654C,2023ApJ...953..150S}. The coronal holes act as magnetic walls pushing the eruptive filaments away, while the others serve as attractors pulling the filaments towards their respective positions. Given the complexity of solar magnetic environment and eruptions, the causes of tilted filament ejections remain difficult to fully explain.

\begin{figure}[ht!]
	\plotone{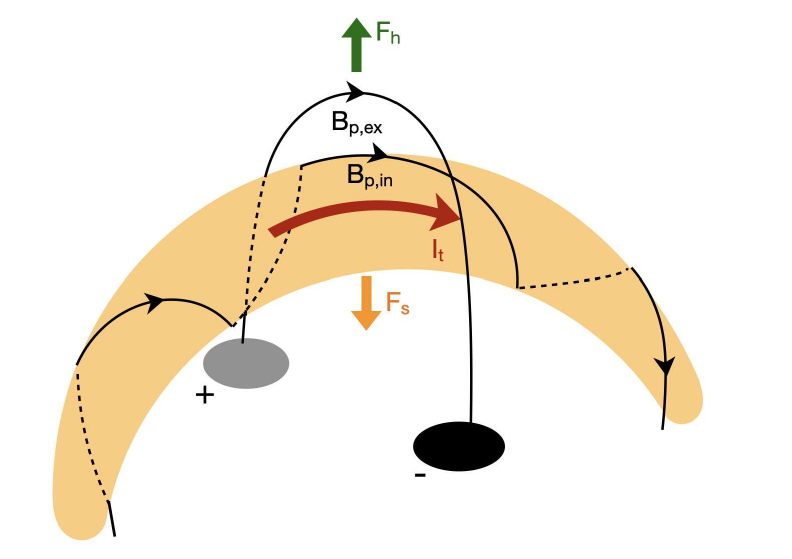}
	\caption{Illustration of the current ring model. The black curves with arrowheads represent the magnetic field lines. The red arrow indicates the direction of the axial current, while the orange and green arrows indicate the directions of the strapping force and hoop force, respectively. The gray and black circles denote positive and negative magnetic polarities, respectively.} \label{fig:cartoon} 
\end{figure}

Following the current ring model \citep{2006PhRvL..96y5002K}, the force balance of the flux rope can be expressed as:
\begin{equation}
	m \frac{d^{2} \textbf{R}}{d t^{2}}=\frac{\mu_{0} I_{\rm t}^{2}}{4 \pi R} (L+ 1 / 2)\hat{\mathbf{r}}+\textbf{I}_{\rm t} \times\textbf{B}_{\rm p,ex}
\end{equation}
where $m$ represents the mass per unit length, $\textbf{R}$ is the major radius of the current ring, $L$ denotes the self-inductance, $\textbf{I}_{\rm t}$ is the axial current and $\textbf{B}_{\rm p,ex}$ means the poloidal component of the external magnetic field. The first term on the right-hand side of the equation is referred to as the self-Lorentz force, also known as the hoop force, while the second term is denoted as the strapping force. As Figure~\ref{fig:cartoon} illustrates, the hoop force always points outward to promote the eruption of the flux rope, whereas the strapping force points downward to hinder the eruption. Since the strapping force originates from the interaction between the ring current and the surrounding magnetic field, the non-uniform magnetic environment could lead to an asymmetric distribution of the strapping force. The region with stronger strapping force would then provide enhanced confinement of filament eruptions. Therefore, is it possible that asymmetric strapping force could also result in the inclination of the filament eruptions?

Because of the line-tying effect, the variation of $\textbf{I}_{\rm t}$ is almost negligible \citep{2006PhRvL..96y5002K}. Hence, $\textbf{B}_{\rm p,ex}$ serves as a good parameter for quantifying the strapping force. Furthermore, how the poloidal field decreases along the ejection distance can be described by the decay index \citep{2010ApJ...718.1388D,2020RAA....20..165L,2024ApJ...964..125S}, which is widely calculated through the horizontal component of potential field and the vertical height \citep{2015SoPh..290.1703M,2015ApJ...814..126Z,2017ApJ...843L...9W}. A more accurate method is to calculate it from the component of the  external magnetic field orthogonal to both the current ring axis and the ejection direction \citep{2019ApJ...870L..21G,2020ApJ...901...13Q,2021ApJ...909...91K,2021NatCo..12.2734Z}. In general, the height where the decay index exceeds 1.5 is treated as the critical height where the torus instability sets in \citep{2006PhRvL..96y5002K,2010ApJ...708..314A,2021MNRAS.501.4703J}. However, the critical value of the decay index is not unique, due to the morphology of the current channels \citep{2010ApJ...718..433O,2010ApJ...718.1388D} and the complex actual situations \citep{2015ApJ...814..126Z,2021ApJ...908...41A}. When the filaments reach the critical height and torus instability takes place, the eruptive filaments start to ascend rapidly \citep{2010ApJ...708..314A,2020ApJ...894...85C}. Moreover, the faster the poloidal field decays, the faster the ejected structure moves in statistics \citep{2012ApJ...761...52X,2017SoPh..292...17D}. These facts imply that the poloidal field distribution might alter the filament ejection trajectory.

This paper aims to figure out how the poloidal field affects the filaments' eruptive behavior and whether there is another reason for the non-radial ejection of the eruptive filaments, by comparing the filaments ejection acceleration and the three-dimensional ejection direction with the poloidal field distribution. Section~\ref{sec:data}~introduces the overview of the selected events, the observed data and the analysis methods. The results are arranged in Section~\ref{sec:result}. Section~\ref{sec:dis_con} shows the discussion and conclusion of this work.

\section{Observations and Methods} \label{sec:data}
For better understanding the relationship between the poloidal field and filaments' kinematics, we choose two eruptive filaments that almost ejected along a straight line without rotation. One of them occurred on 2011 March 7 (hereafter referred to as Event 1), originating from active region (AR) 11164 and consequently generating an M3.7 flare at N24$^\circ$ W60$^\circ$. This flare started at 19:43 UT, peaked at 20:12 UT and ended at 20:58 UT. The other filament erupted on 2014 February 25 from AR 11990 at S12$^\circ$ E77$^\circ$, resulting in an X4.9 flare which initiated at 00:39 UT, reached its peak at 00:49 UT and ended at 01:03 UT (hereafter, Event 2). Both events were accompanied by subsequent CMEs, revealing that both filaments ejected from the Sun successfully. Moreover, both events were almost simultaneously observed by the Solar Dynamics Observatory (SDO; \citealt{2012SoPh..275....3P}) and Solar TErrestrial RElations Observatory (STEREO; \citealt{2008SSRv..136....5K}) from two separate perspectives, conducive to constructing the three-dimensional ejection paths of the two filaments using the triangulation method. 

The Atmospheric Imaging Assembly (AIA; \citealt{2012SoPh..275...17L}) aboard SDO includes two ultraviolet (UV) channels and seven extreme ultraviolet (EUV) channels. All channels have the same spatial resolution of 0.6 arcsec pixel$^{-1}$ but different cadence of 24 s and 12 s for UV and EUV wavebands. The hot wavebands, such as 94~\AA, are able to well exhibit the activities in the corona, while the cold wavebands like 304~\AA~and 1600~\AA~have good responses to the phenomena in lower atmosphere. The STEREO mission comprises two identical spacecraft, named STEREO-A and STEREO-B, respectively, one orbiting ahead of the Earth and the other behind. These two spacecraft can also capture the full-disk images at 304~\AA~with the pixel size of about 1.6 arcsec by Extreme-Ultraviolet Imager (EUVI; \citealt{2004SPIE.5171..111W}). The same wavelength observations but different orbiting speeds prompt the synchronous observations from different perspectives, making the three-dimensional location measurement based on the tie-pointing technique possible (e.g. \citealt{2011ApJ...730..104J, 2011ApJ...739...43L,2020ApJ...892...54X,2023ApJ...953..150S}). The tie-pointing technique establishes correspondences between identical features visible in the 2D observations from different views and reconstructs 3D geometry by triangulating the object points using known telescope positions \citep{2006astro.ph.12649I}. For Event 1, STEREO-A was approximately 88$^\circ$ ahead the Earth on 2011 March 7. As for Event 2, the separation between the STEREO-B  and the Earth was roughly 160$^\circ$ on 2014 February 25. We adopted the \textit{scc$\_$measure} program in the Solar SoftWare (SSW) library to measure the coordinates of the same features in simultaneous observations from two separated spacecrafts. This program displays near-simultaneous multi-perspective images. Upon selecting an object point along a filament spine in one image, the routine plots the corresponding epipolar line which connects the object point and the first perspective's optical center on the other image, as indicated by the dashed lines in Figures~\ref{fig:tria_20110307}~and~\ref{fig:tria_20140225}. By clicking the intersection of the filament spine and this epipolar line, the program automatically computes the 3D coordinates of the selected object point. Note the measured coordinates are based on the Stonyhurst heliographic coordinates whose origin is the intersection point of the solar equator and the central meridian as viewed from Earth.

\begin{figure}[ht!]
	\plotone{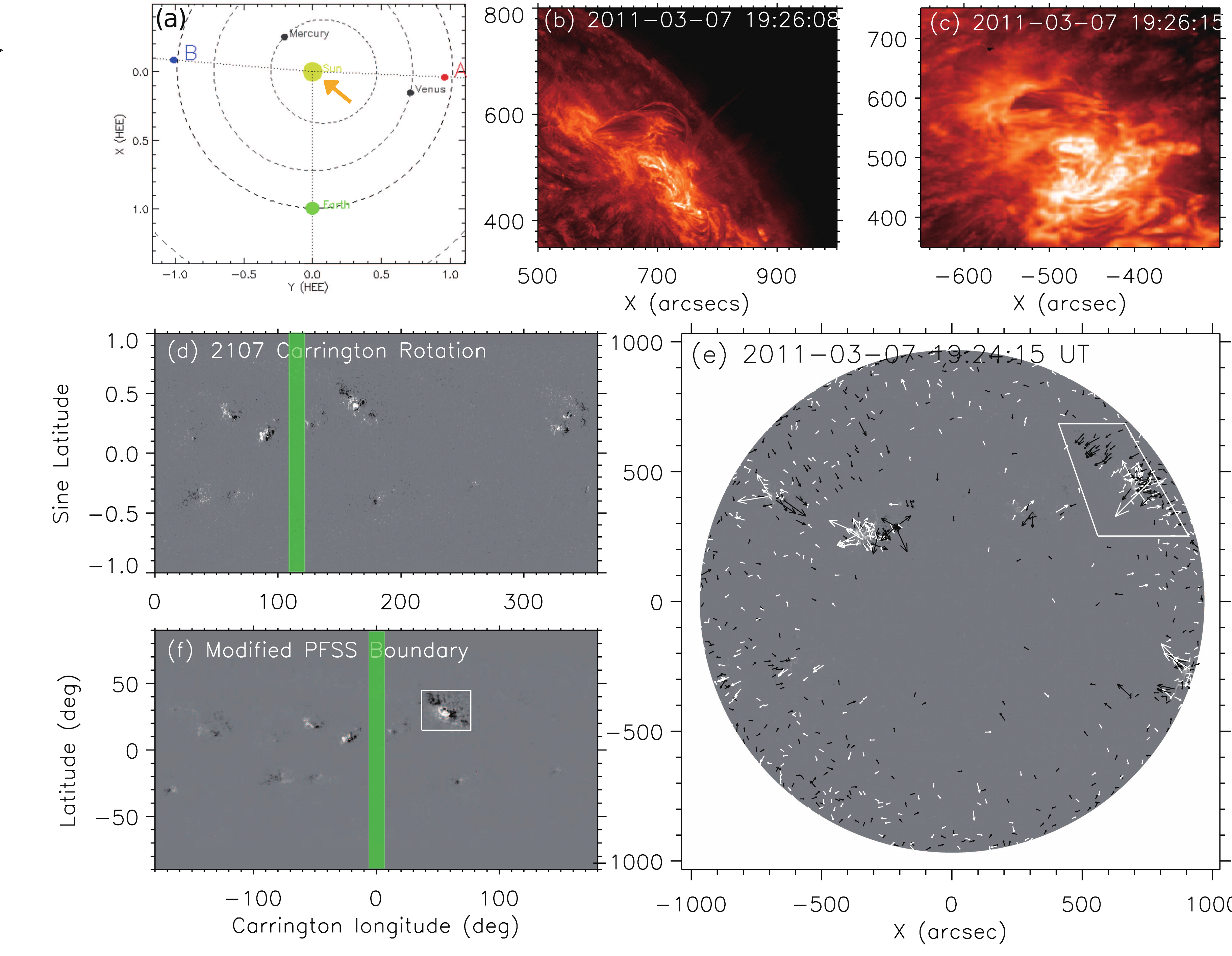}
	\caption{Overview of the eruptive filament on 2011 March 7. (a) Locations of STEREO satellites at 19:30 UT. The orange arrow points to the eruptive filament of interest. (b) The filament image observed by SDO/AIA 304~\AA~channel at 19:26:08 UT. (c) The image of the same filament observed by STEREO-A in 304~\AA~waveband at almost the same time. (d) The synoptic magnetic field of 2107 Carrington Rotation, with the green-shaded band marking the central meridian at 19:00 UT.  (e) Photospheric vector magnetogram captured by SDO/HMI at 19:24:15 UT. The white quadrilateral encircles the area designated for replacement. (f) Modified synoptic map for PFSS extrapolation. The green shadow shows the shifted central meridian and the white rectangle marks the corresponding replaced area.} \label{fig:mag_20110307} 
\end{figure}

\begin{figure}[ht!]
	\plotone{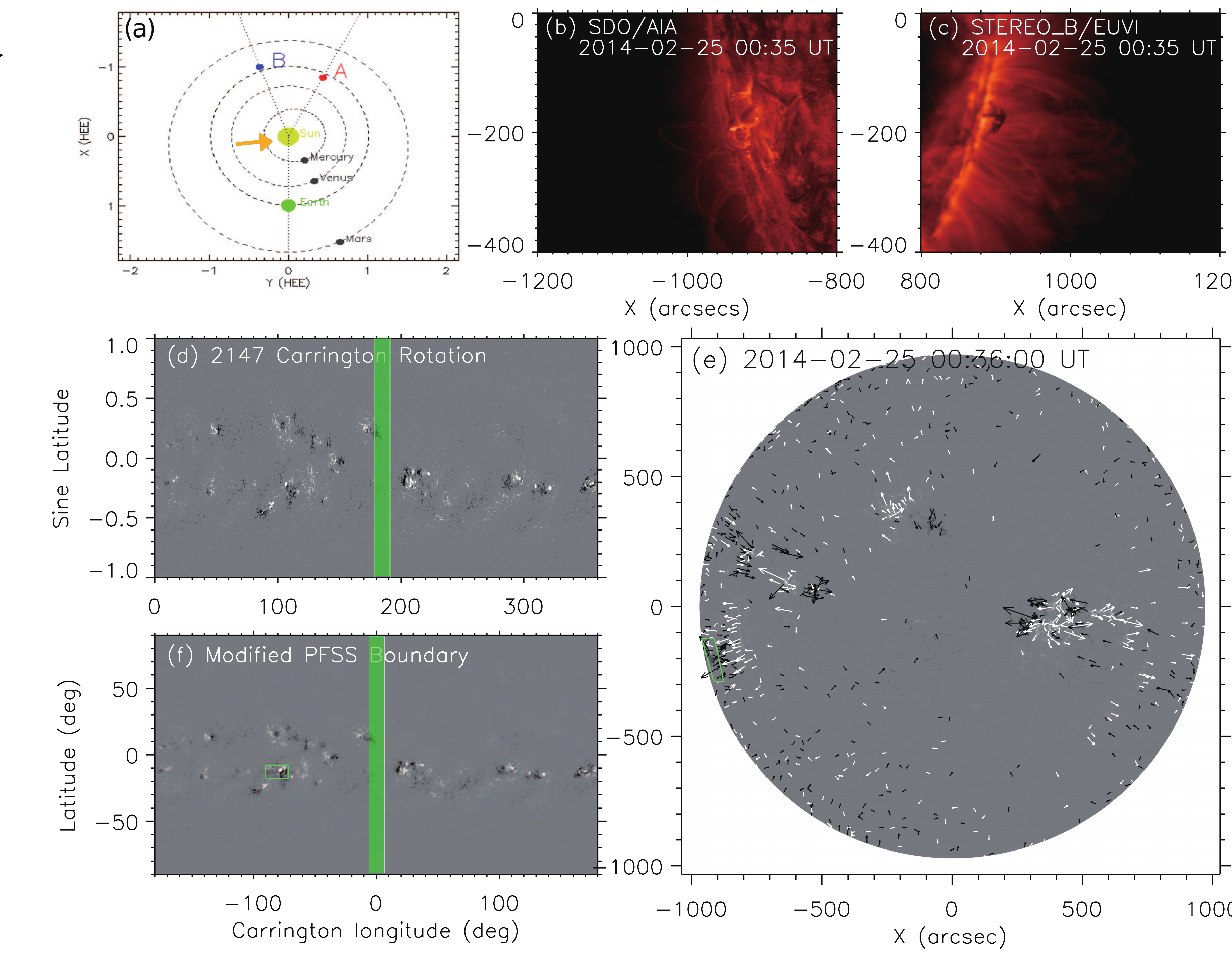}
	\caption{Overview of the filament eruption on 2014 February 25. (a) Orbital positions of STEREO satellites at 00:35 UT. The orange arrow indicates the location of the eruptive filament. (b) Pre-eruptive filament structure observed by SDO/AIA 304 Å channel at 00:35:31 UT. (c) Quasi-simultaneous 304 Å observation of the same filament from STEREO-B. (d) Synoptic magnetic field map for Carrington Rotation 2147, with the green-shaded band designating the central meridian at 00:35 UT.  (e) Photospheric vector magnetogram acquired by SDO/HMI at 00:36:00 UT. The green quadrilateral at the eastern limb demarcates the target region for replacement. (f) Modified synoptic map for Potential Field Source Surface (PFSS) extrapolation, where the green band indicates the shifted central meridian and the rectangle highlights the corresponding replaced area.} \label{fig:mag_20140225}
\end{figure}

To acquire the large-scale three-dimensional coronal magnetic field covering the erupted filaments, we employ  the Potential Field Source Surface (PFSS; \citealt{1969SoPh....6..442S,1969SoPh....9..131A,2014ApJS..214....4P}) method implemented in the Message Passing Interface Adaptive Mesh Refinement Versatile Advection Code (MPI-AMRVAC; \citealt{2014ApJS..214....4P,2018ApJS..234...30X,2023A&A...673A..66K}).  This method can extrapolate the magnetic field from 1 to 2.5 solar radii, importing the global $B_{\rm r}$ map as the calculation boundary. In this work, the boundary is assigned to the synoptic data \citep{2018arXiv180104265S}  observed by Helioseismic and Magnetic Imager (HMI; \citealt{2012SoPh..275..229S}) on SDO, named as \textit{hmi.synoptic\_mr\_polfil}. The synoptic map consists of a series of $B_{\rm r}$ magnetogram within 2.2$^\circ$ longitude around the solar central meridian over one Carrington rotation. Therefore, the  $B_{\rm r}$ values in the active regions we are interested in might be observed 3--4 days before or after the time when the filaments erupted, which might be inaccurate for extropolation because of the rapid evolution of the magnetic field in active regions. To correct the differences caused by this time lag, we extract the magnetic field in the region of interest from the full-disk photospheric vector magnetograms acquired half hour preceding the eruptions (Figures~\ref{fig:mag_20110307}(e) and \ref{fig:mag_20140225}(e)), calculate the radial magnetic field strength, and replace $B_{\rm r}$ in the corresponding region in the synoptic map (Figures~\ref{fig:mag_20110307}(f) and~\ref{fig:mag_20140225}(f)). Meanwhile, the $y$-axis of the synoptic map represents the sine value of latitude, and the central meridian at the observation time may not coincide with the 0° longitude because the Carrington longitude is inversely proportional to time. Prior to application, we use the bilinear interpolation to convert the $y$-axis to latitude and rotate the central meridian of the observation day to the 0° longitude line (Figures~\ref{fig:mag_20110307}(f) and~\ref{fig:mag_20140225}(f)). This ensures that the coordinate system of the extrapolated global coronal potential field is consistent with the coordinate system on the day of the filament eruptions.

\section{Results} \label{sec:result}
\subsection{Three-dimensional trajectory and axial current direction of the erupted filaments}\label{subsec:3d_path}
\begin{figure}[ht!]
	\plotone{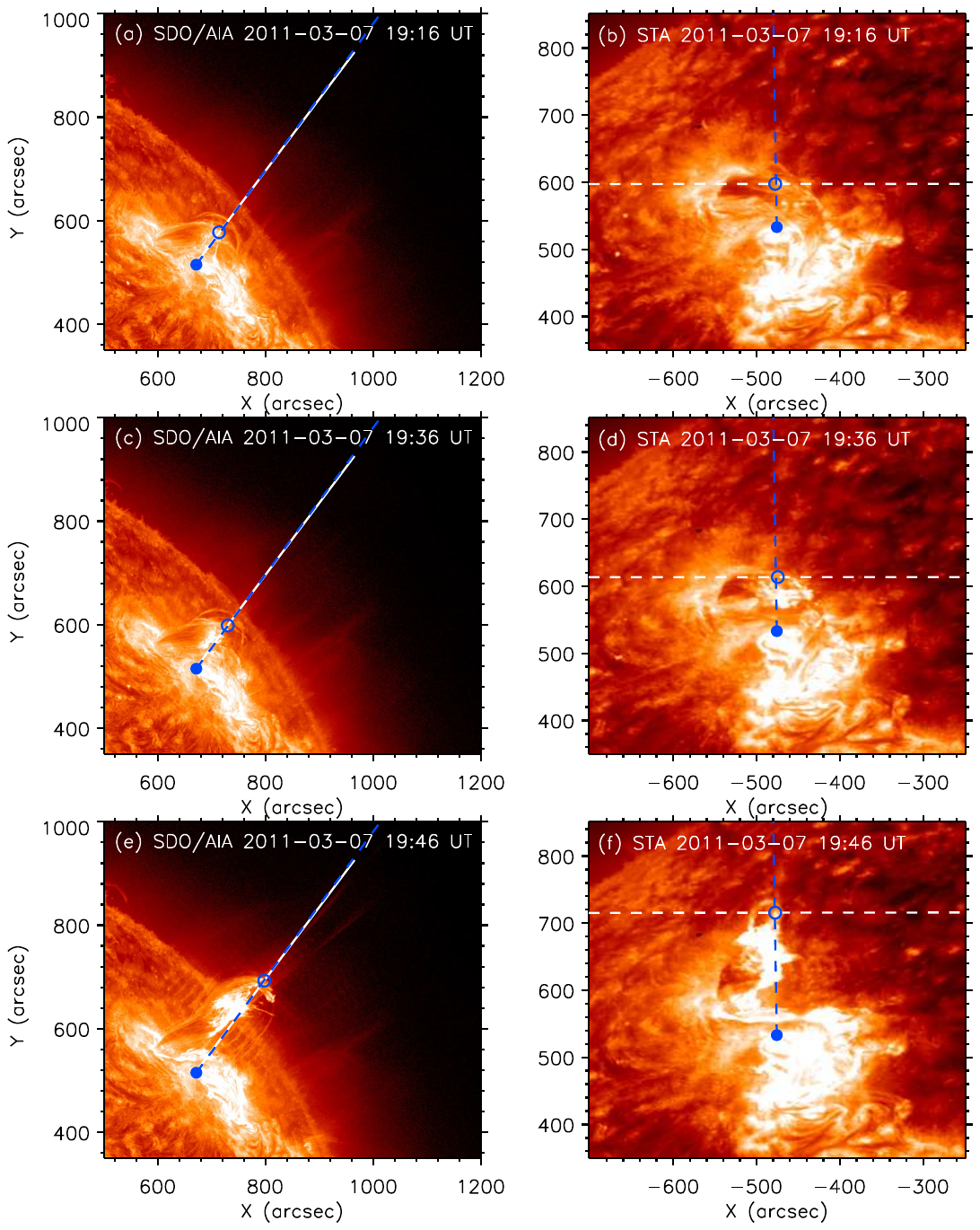}
	\caption{Triangulation of the vertices of the eruptive filament at three discrete time points on 2011 March 7. The white solid lines delineate the slice used to trace the ejection trajectory of the filament in the plane of sky observed by SDO/AIA. The white dashed lines represent the epipolar lines aligned with the SDO/AIA line of sight. The blue circles indicate the measured points under the 3D reconstruction. The blue dashed lines depict ejection trajectories derived from 3D linear fitting, projected onto the planes of the sky for SDO and STEREO observations, respectively. The blue dots signify the intersections of the ejection paths with the photosphere. A 10-second animation exhibiting the full evolution of the eruption in 304~\AA~by AIA is available online.
		} \label{fig:tria_20110307}
\end{figure}

\begin{figure}[ht!]
	\centering
	\includegraphics[scale=0.8]{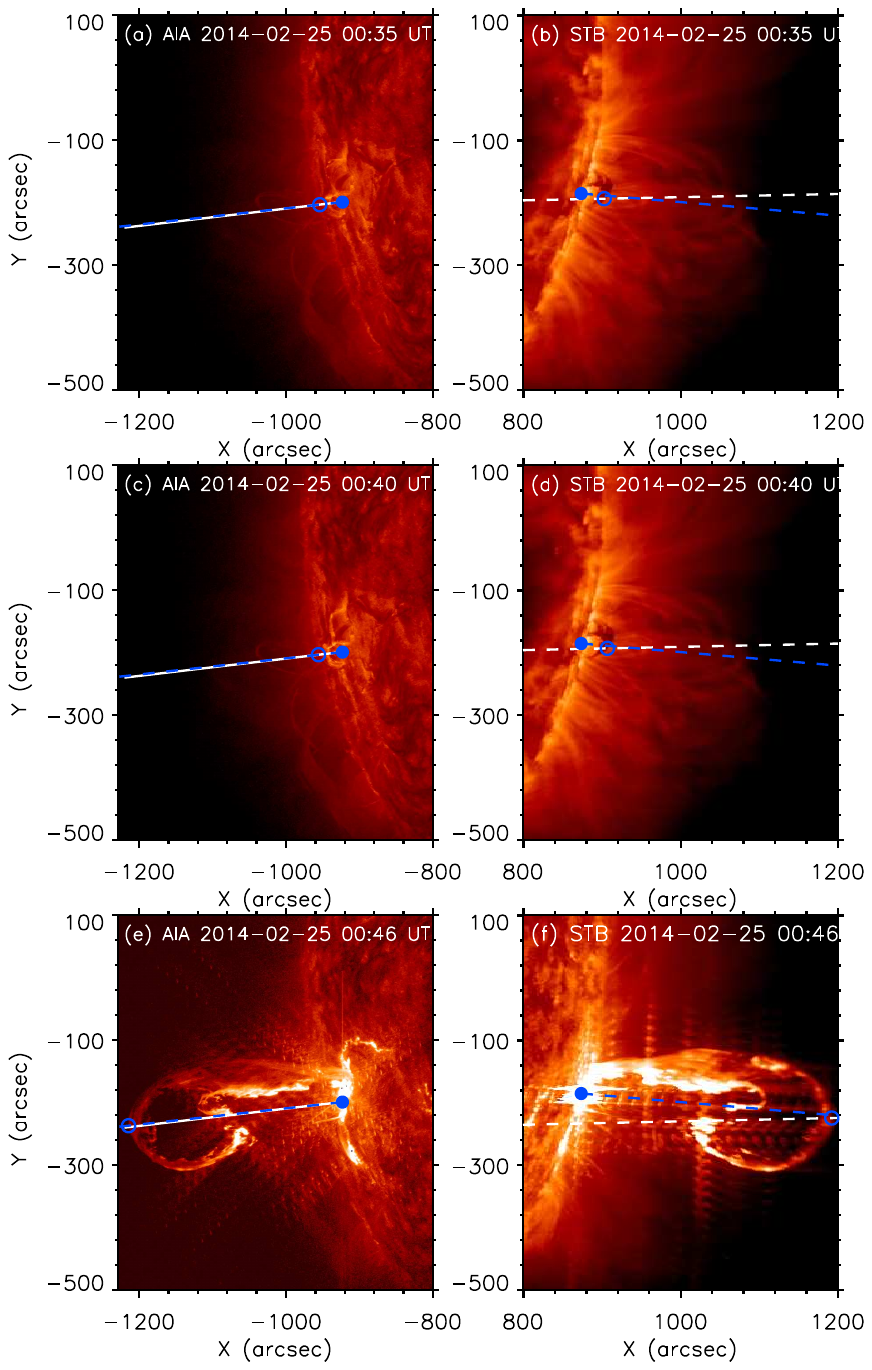}
	\caption{Triangulation of the eruptive filament vertices at three different moments on 2014 February 25. The legend follows the same conventions as Figure~\ref{fig:tria_20110307}, where white solid lines indicate the slices for tracking the filament ejection, white dashed lines denote the corresponding epipolar lines, and blue circles represent triangulated measurement points. The blue dashed lines delineate the projected trajectories of the fitted ejection paths, with the blue dots showing their intersections with the photosphere. The online supplemental material includes a 10-second animation demonstrating the complete eruption process observed by AIA in 304~\AA~ waveband.} \label{fig:tria_20140225}
\end{figure}

To track the trajectory of the filament eruptions in AIA 304~\AA~images, we make a slice along the propagation direction of the filament apexes (white solid lines in Figures~\ref{fig:tria_20110307} and~\ref{fig:tria_20140225}). Subsequently, we mark the slice location in the AIA images and use the tie-pointing method to measure the three-dimensional coordinates of the filament apexes through the dual-viewpoint observations from SDO/AIA and STEREO/EUVI at different times (blue circles in Figures~\ref{fig:tria_20110307} and~\ref{fig:tria_20140225}). By performing the three-dimensional linear fitting method on these sampling points, we further acquire the three-dimensional eruption path of the filaments and calculate the ejection inclination angle, which is the angle between the ejection path and the radial direction. The inclination angles of Event 1 and Event 2 are 14.8$^\circ$ and 26.4$^\circ$, respectively. We project the fitted three-dimensional path back onto the SDO/AIA and STEREO/EUVI planes of sky (blue dashed line in Figures~\ref{fig:tria_20110307} and~\ref{fig:tria_20140225}), and compare it with the original slices in the AIA images. The nearly perfect coincidence suggests that the measurement of the three-dimensional paths is relatively accurate.

\begin{figure}[ht!]
	\plotone{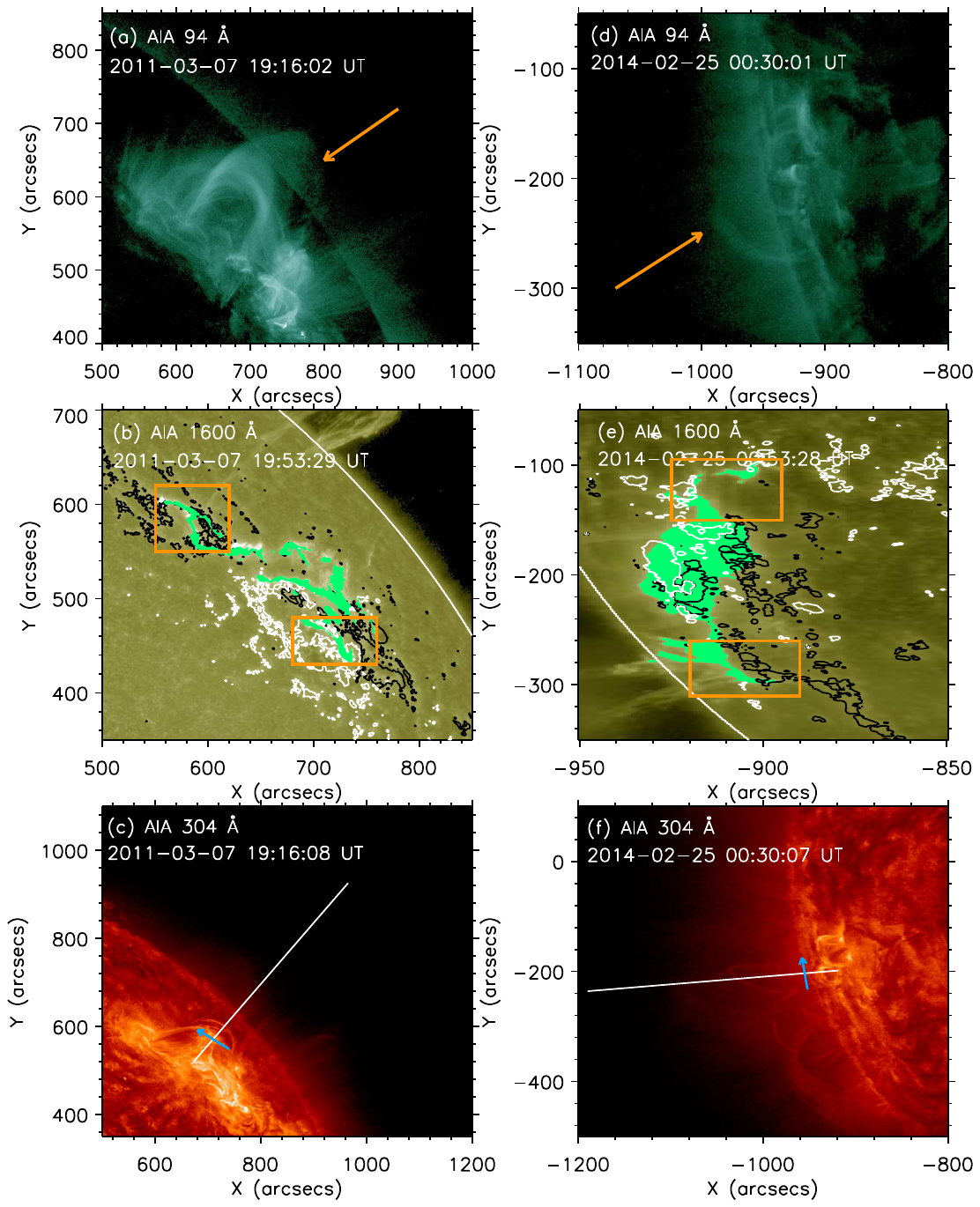}
	\caption{(a, d) AIA 94~\AA~images showing the pre-existing hot channels, indicated by the orange arrows. (b, e) AIA 1600~\AA~images, with the green masks outlining flare ribbons. The white and black lines sketch the contours of 400 G and -400 G radial magnetic field, respectively. The orange rectangles mark the elongated hook brightenings of the flare ribbons. (c, f) AIA 304~\AA~images superimposed with the directions of the axial currents within the magnetic flux rope, as indicated by the blue arrows. } \label{fig:current}
\end{figure}

As indicated by the orange arrows in Figures~\ref{fig:current}(a) and (d), pre-existing hot channels are visible in the 94 Å images for both events. Additionally, there are hook-shaped brightenings elongating from the ends of the flare ribbons in the 1600 Å images (outlined by yellow rectangles in Figures~\ref{fig:current}(b) and (e)). Besides, the locations of these hook-shaped brightenings coincide with the fallback positions of the filament material. These observed features imply that both filaments might be supported by magnetic flux ropes (e.g. \citealt{2012NatCo...3..747Z,2019A&A...621A..72A,2020Innov...100059X,2022ApJ...940...62L,2024RvMPP...8...27S,2024A&A...682A...3X}). This inference is consistent with the results reported by \cite{2013ApJ...763...43C} and \cite{2014ApJ...797L..15C}. Furthermore, combining the photospheric radial magnetic field, we can infer that the filament on 2011 March 7 is sinistral (positive helicity) while the filament on 2014 February 25 is dextral (negative helicity), according to a filament chirality rule that was first proposed by \cite{2010ApJ...714..343G} and later visualized by \cite{2014ApJ...784...50C}.  This means that the axial current of Event 1 flows from the positive polarity to the negative polarity while the axial current of Event 2 flows in a reverse way. Both filaments violate the hemispheric helicity rule \citep{2003ApJ...595..500P}, implying them prone to eruptions.

On the basis of the definition, the poloidal component of the magnetic field is perpendicular to both the ejection direction and the axial current direction. However, we only confirm how the axial current flows between the polarities, but we still do not know the direction of the axial current at the filament apex. To obtain the poloidal magnetic field distribution along the ejection path of the filaments, we also employ the tie-pointing method to measure five consecutive points on both sides of the filament apex at a certain moment before the eruption. The five points are also fitted by three-dimensional linear fitting method to determine the filament spine at the apex. Combining the spine location with the previously deduced flow direction of the axial current, we are able to finally obtain the direction of the axial current at the apex (blue arrows in Figures~\ref{fig:current}(c) and (f)).

\subsection{Relationship between the acceleration of filament eruptions and the poloidal magnetic field}\label{subsec:acc}

\begin{figure}[ht!]
	\plotone{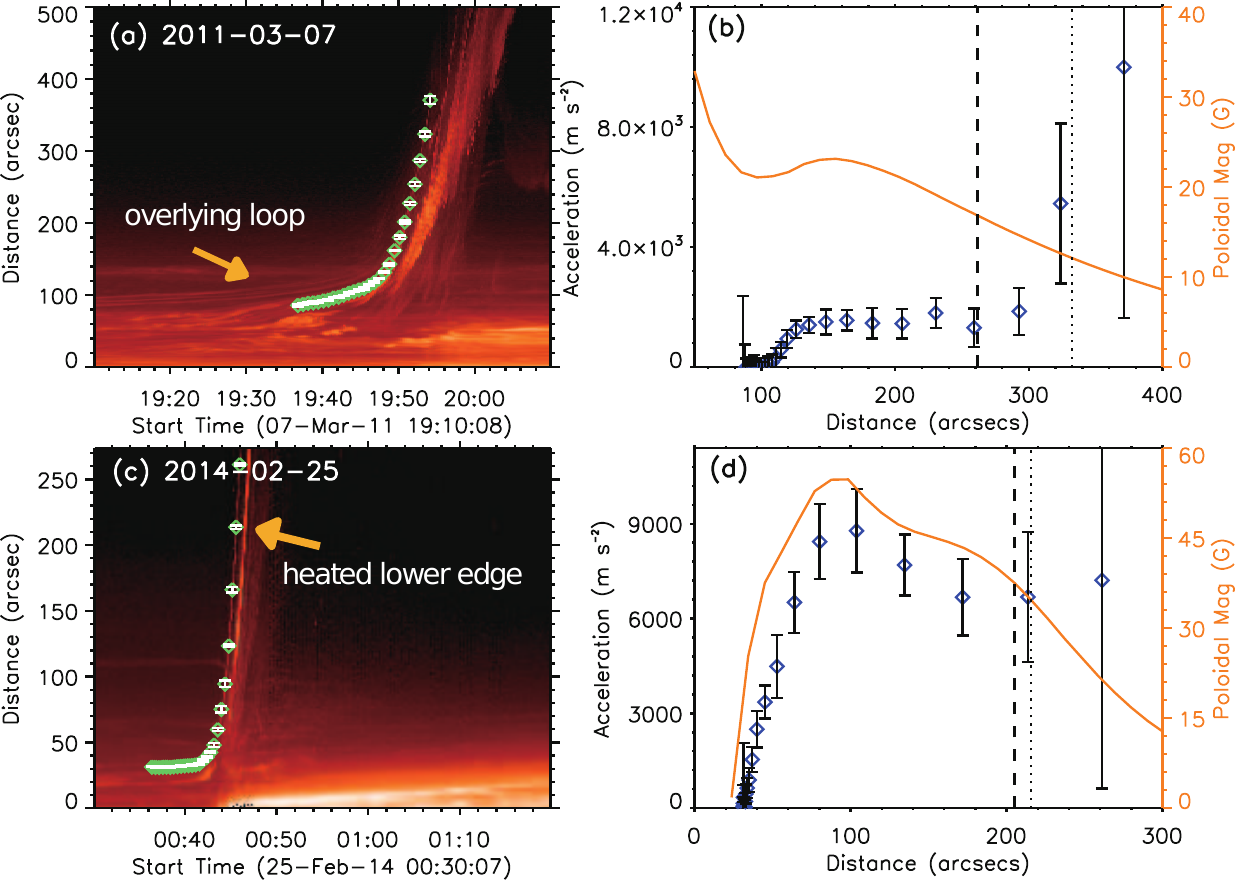}
	\caption{(a) Time-distance map of the filament erupting on 2011 March 7. The green diamonds with white error bars trace the filament propagation. (b) Comparison between the acceleration and poloidal field, which are shown by the blue diamonds with black error bars and the orange solid line, respectively. The vertical dashed line marks the distance where the decay index just exceeds 1 and the vertical dotted line indicates where the decay index reaches 1.5. (c) and (d) exhibit the same results of the eruptive filament on 2014 February 25.} \label{fig:acc}
\end{figure}

As shown by the time-distance diagrams (Figures~\ref{fig:acc}(a) and (c)), both filaments experience a rapid rise phase during the ejection process. We measure the ejection track of the filaments on the time-distance diagrams at fixed time intervals and repeat the measurement five times. The average of these five measurements is regarded as the ejection profile of the filament, and the standard deviation is deemed to be the measurement uncertainty. By taking the second derivative of the ejection distance with respect to time, we obtain the ejection acceleration of the filaments, corresponding to the blue diamonds in Figures~\ref{fig:acc}(b) and (d). It is obvious, in both events, that the acceleration increases monotonically at the beginning, then levels off and even begins to decrease after reaching a certain altitude. Travelling a longer distance, the acceleration shows a tendency to ascend again.

To investigate the reasons for the observed changes in the ejection acceleration of both filaments, we calculate the poloidal magnetic field distribution along the three-dimensional ejection paths by  $B_{\rm pol} = \textbf{e}_{\rm{pol}}\cdot\textbf{B}_{\rm p}$, where $\textbf{e}_{\rm pol}~\times~(\textbf{e}_{\rm axis}~\times~\textbf{e}_{\rm ejec})= 0$  and  ${\textbf{B}}_{\rm p}$ represents the potential field \citep{2019ApJ...870L..21G}. For comparison, we further acquire the poloidal field distribution with the ejection distance projected back onto the AIA plane of sky (orange solid lines in Figures~\ref{fig:acc}(b) and (d)) and compare it with the acceleration. It is not difficult to observe that the ejection acceleration of both events increases sharply at the beginning, perhaps under the effect of magnetic reconnection. When the poloidal magnetic field strengthens to a certain value, the acceleration begins to be suppressed (Event 1) or even decreases (Event 2). Subsequently, the poloidal field starts to decay again. And as the eruptive filaments reach the height where the decay index just surpasses 1, the acceleration begins to gradually increase once more. This may suggest that the variation trend of the ejection acceleration has a strong correlation with the strength of the poloidal field.

\begin{figure}[ht!]
	\plotone{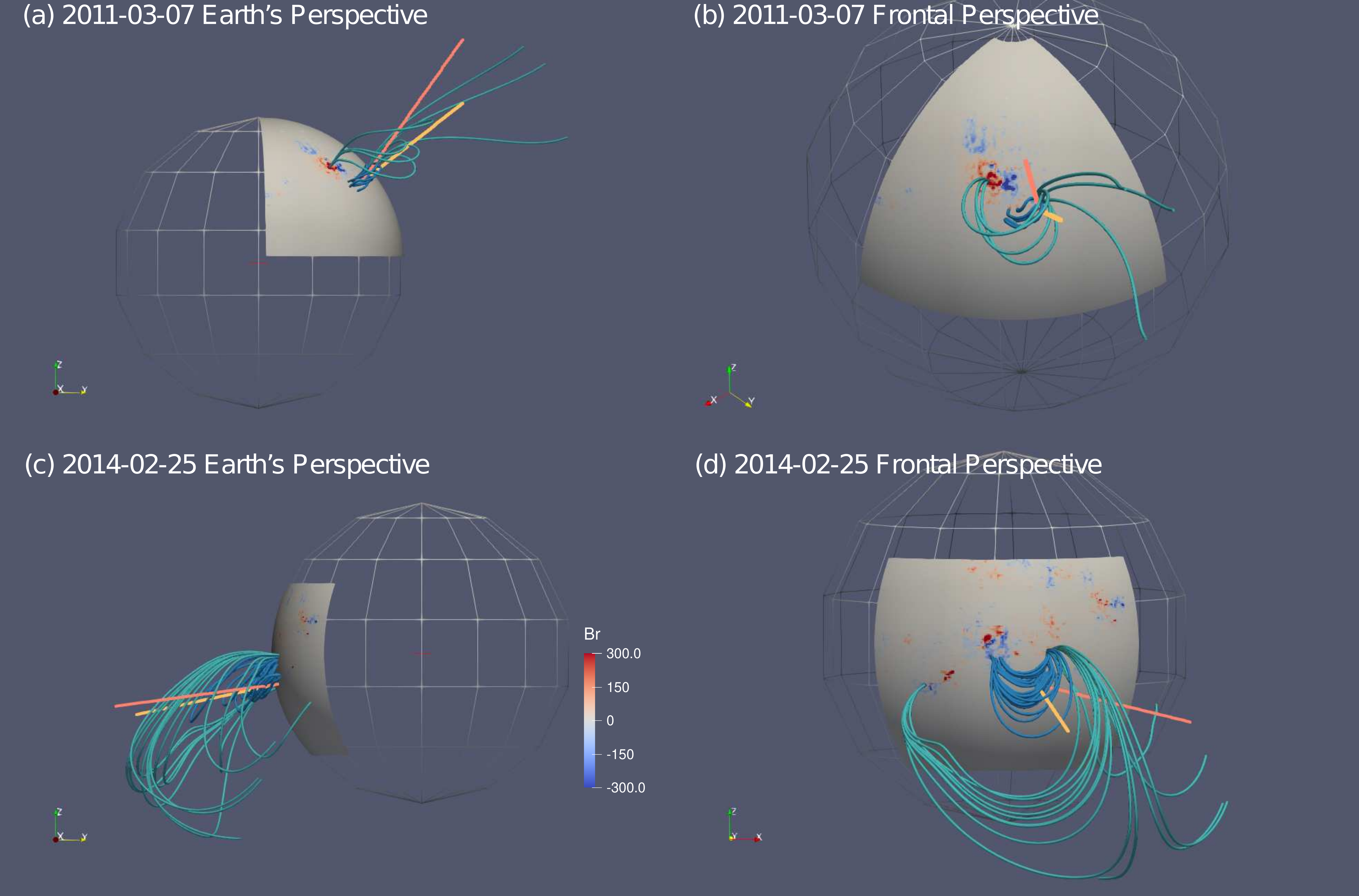}
	\caption{Extrapolated magnetic field lines enveloping two eruptive filaments computed by the PFSS model. The dark blue lines indicate the lower magnetic field lines while the cyan lines represent higher. The orange straight lines denote the actual ejection path of two filament eruptions, with the yellow lines marking the radial direction for comparison. (a, b) PFSS model of Event 1 from Earth's and frontal views. (c, d) PFSS model of Event 2 from the same perspectives.  } \label{fig:field}
\end{figure}

To figure out why the poloidal field changes, we exhibit the overlying magnetic field of two filaments from two different perspectives in Figure~\ref{fig:field}.  Both events pass through two sets of loop systems, consisting of the lower bipolar loop system and upper quasi-bipolar loop system. For Event 1, the bipole generating the lower loop system is relatively simple and unchanged, so the poloidal field monotonously decreases at the beginning. With the joining of the upper loop system, the poloidal field increases and shows a peak due to the subsequent decrease. Variously, the poloidal field of Event 2 increases at first because the lower traverse loops originate from the region gradually approaching the core of the negative polarity of AR 11990. Then, the lower loop system expands too fast with height, causing the decrease of the poloidal field. The joining of the upper loop system results in the bump at the distance of around 180$''$ .

\subsection{Causes of the non-radial ejections}\label{subsec:tilted_eject}
From the three-dimensional linear fitting of the ejection path, we know that both filaments are inclined with respect with the radial direction. The inclination angles of these two filaments are 14.8° and 26.4°, respectively. Figure~\ref{fig:field} shows that the filaments seem to be ejected near the location where the upper loop system splits into two parts, one of them presenting as open field. This might be because the open magnetic field lines at that location are approximately vertically oriented, with relatively smaller horizontal components, resulting in weaker interception on the flux rope and the filament.

\begin{figure}[ht!]
	\plotone{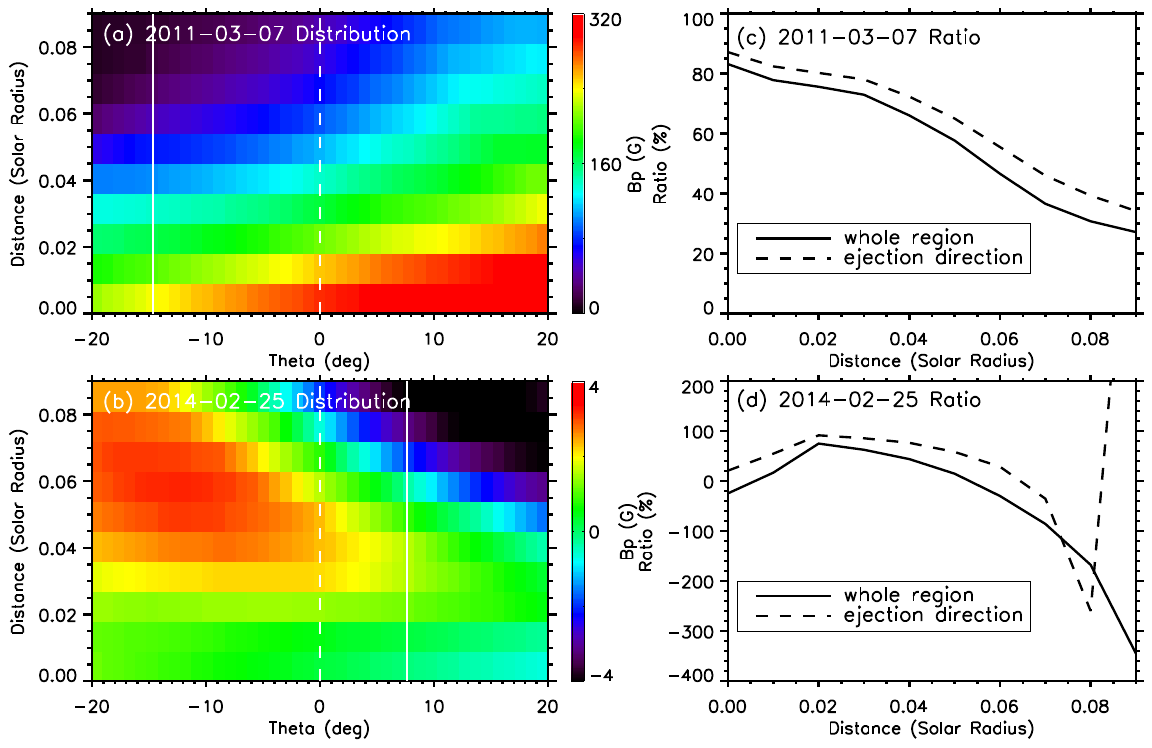}
	\caption{(a, b) Two-dimensional distributions of the poloidal field above the filaments. The dashed line means the projection of the radial direction onto the calculated plane, while the solid line indicates the direction of the realistic ejection path projected onto the calculation plane. (c, d) Evolution of the ratio of the averaged poloidal field strength between the ejection-direction side and the opposite side with height. Solid lines denote the ratio of the averaged strength from the radial direction to the computational boundary (±20°), while dashed lines represent the ratio of the averaged strength from the radial direction to the projection of the ejection direction.} \label{fig:poloidal_field}
\end{figure}

To further investigate the reason for the inclined ejection of the filaments, we calculate the spatial distribution of the poloidal component of the magnetic field along different directions. To achieve this, we measure the apex of the filaments five times based on the dual-perspective observations at the moment before eruptions, and take the average of these five measurements as the starting point of the filament ejections. We then set the axial current direction determined in Section~\ref{subsec:3d_path}~at the apex as the normal vector, and calculate the interception plane passing through the starting point. Note that this interception plane may not pass through the solar center. We compute the poloidal field distribution along different directions from the starting point in the interception plane (Figures~\ref{fig:poloidal_field}(a) and (b)). Because the ejection directions are derived from the 3D trajectory of the apexes of the filaments, the measured ejection directions are not perpendicular to the directions of the axial current. We project the observed ejection direction and the radial direction onto the calculation plane for comparing (see solid and dashed lines in Figures~\ref{fig:poloidal_field}(a) and (b)). The result reveals that the poloidal field above both filaments is stronger on one side of the radial direction and weaker on the other side, with a monotonically decreasing trend in distribution with respect to angle. Furthermore, the actual ejection directions of the filaments both incline towards the weaker side, indicating that when the distribution of the poloidal field is asymmetric, the filaments will be ejected towards the side with weaker confinement.

 Furthermore, we calculate the ratio of the averaged poloidal magnetic field strength on the ejection side to that on the symmetrically opposite side relative to the radial direction, which is displayed in Figures~\ref{fig:poloidal_field}(c) and (d). The solid lines denote the ratio averaged over whole calculation regions shown in Figures~\ref{fig:poloidal_field}(a) and (b), labeled as $r_{\rm all}$, while the dashed lines represent the ratio averaged within the projected range of the ejection direction, labeled as $r_{\rm eje}$. For both events, $r_{\rm all}$ is notably smaller than $r_{\rm eje}$. For Event 1, $r_{\rm all}$ decreases gradually from 83\%~to 27\%~as height increases, while $r_{\rm eje}$ monotonically declines from 87\%~to 34\%. For Event 2, both $r_{\rm all}$ and $r_{\rm eje}$ exhibit an initial increase followed by a decrease, reaching local maxima of 75\%~and 92\%, respectively. Subsequently, they decline to negative values, indicating a reversal of the poloidal magnetic field at greater heights. This reversal implies that the poloidal field no longer suppresses the eruption but instead facilitates it (the final positive value of $r_{\rm eje}$ arises because the poloidal field reverses on both sides). These further demonstrate that the strapping force on the ejection side is weaker than on the opposite side.

\section{Discussion and Conclusion}\label{sec:dis_con}
In this paper, we analyze the relationship of the kinematics and the overlying poloidal field of two filament eruptions. One occurred in AR 11164 on 2011 March 7 (Event 1), and the other erupted in AR 11990 on 2014 February 25 (Event 2). The supporting structures of both filaments are magnetic flux ropes, as deduced by the pre-existing hot channel structures and the elongated hook-shaped brightennings. According to the helicity judgement approach, Event 1 has a positive helicity in the Northern Hemisphere, whereas Event 2 possesses a negative helicity in the Southern Hemisphere. Thus, neither event satisfies the hemispheric rule and is prone to eruption.  We utilize the tie-pointing technique and three-dimensional linear fitting to measure the three-dimensional paths of the filament eruptions and the current direction at the apex of the filaments at the initial moment. Besides, we reconstruct the global coronal magnetic field on the observed days using the PFSS model with the modified boundary, and then calculate the poloidal magnetic field distribution along the eruption paths of the eruptive filaments and within the interception plane perpendicular to their axes.

The profile of the acceleration reveals that the acceleration of both filaments increases initially, subsequently levels off and finally ascends again. To figure out the cause of this change, we compare the acceleration of the filament eruptions with the poloidal field along the eruption paths. It shows that the increase and decrease of the acceleration are closely related to the decay and enhancement of the poloidal field. At first, the acceleration of both events grows rapidly, possibly resulting from the underneath magnetic reconnection. But when two filaments reach the location where the poloidal field appears a small peak (Event 1) or increases to a certain value (Event 2), the strapping force of the background magnetic field becomes strong enough to suppress the ejection of the flux ropes,  causing the acceleration of the flux ropes to stop growing (Event 1) or even start decreasing (Event 2). The local strengthening of the poloidal field originates from the joining of the additional loop system. After that, the poloidal field decays with distance monotonously. When passing through the location where the decay index exceeds 1, which is a typical critical value for a straight current tube \citep{2010ApJ...718.1388D}, the ejection acceleration rises again. This might result from the fact that during the ejection of the flux rope, the rope  apex gradually stretches and aligns into a straighter configuration. This means that when the flux rope reaches the critical height of a straight current tube, the downward strapping force will decline faster than the upward hoop force, leading to the second increase in the ejection acceleration of the flux rope. 

Based on the results of the dual-perspective measurements and three-dimensional linear fitting, we are able to confirm that the inclination of the ejection is 14.8° for Event 1 and 26.4° for Event 2. Both events are non-radial ejections. To investigate the reason for the inclined filament ejections, we calculate the poloidal field distribution at different angles and heights from the initial filament apex on the interception plane. The calculation shows that the poloidal field is asymmetric, exhibiting a monotonic decrease or increase as the angle changes. In addition, we compare the two-dimensional poloidal field distribution with the projected ejection direction and calculate the percentage of the spatial-averaged poloidal field strength on the ejection side to that on the symmetrically opposite side. The comparison reveals that the eruptive filaments propagate along the side with a weaker poloidal field, where the strapping force of the background magnetic field on the flux rope is relatively weak.

Non-radial ejection of filaments is quite common \citep{2005ApJ...628L.163W,2013SoPh..287..391P,2013ApJ...773..162B,2021A&A...647A..85D,2023ApJ...953..150S}, but the explanation for it remains inconclusive. The lateral magnetic reconnection caused by newly emerging flux could generate a region with lower magnetic pressure,  leading to the non-radial ejection \citep{2000ApJ...545..524C,2005ApJ...628L.163W,2022ApJ...933..148C}. The asymmetrical magnetic reconnection might also account for the deflection by the slingshot effect of horizontally curved magnetic field lines, which provide a strong horizontal magnetic tension force \citep{2013ApJ...779..129K}. In the simulations of \cite{2024MNRAS.533L..25L}, the asymmetrical magnetic reconnection could result from the asymmetrical photospheric flux, but the eruption intensity decreases with the increase of the asymmetry, which means that the ejection with large inclination angle might be hard to occur. Besides the magnetic reconnection, the deflection caused by the ambient magnetic environments has been proven by numerous observations and simulations \citep{2010ApJ...708..314A,2013SoPh..287..391P,2020AdSpR..65.1654C,2020SoPh..295..126S,2023ApJ...953..150S}, constitutionally due to the asymmetry of magnetic pressure. Considering that the ejection of filaments can be hindered and confined by the overlying magnetic field \citep{2010ApJ...725L..38G,2014ApJ...787...11J}, our work proves that the asymmetry of the confinement might result in non-radial ejection, as well. Similar work has been done by \cite{2024ApJ...974..168K}. But they compared the ejection direction with the distribution of decay index, which expresses the decrease rate of the poloidal field with height, and discussed that the flux rope was more torus-unstable towards the direction with decay index exceeding the threshold. We compare the ejection direction with the distribution of the poloidal field, which represents the distribution of strapping force. This comparison directly displays how the confinement of the background magnetic field leads to the inclined ejection of the eruptive filaments. However, it needs further investigations how to quantify the asymmetry of the confinement of the background magnetic field and how to use the quantitative value to predict the accurate ejection direction of the eruptive filaments.

\begin{acknowledgments}
Thanks a lot to the anonymous reviewer for the constructive and valuable suggestions. SDO is a mission under NASA's Living With a Star Program, while STEREO is the third mission of NASA's Solar Terrestrial Probes program. We are grateful for the easy access to the calibrated data facilitated by the SDO and STEREO teams. This study is supported by the National Key R\&D Program of China (2022YFF0503004, 2021YFA1600504, and 2020YFC2201201) and NSFC (12333009).

\end{acknowledgments}


\bibliography{manuscript}{}
\bibliographystyle{aasjournal}

\end{document}